\documentclass[11pt]{iopart}

\newcommand{\bea}{\begin{eqnarray}}
\newcommand{\eea}{\end{eqnarray}}

\begin{document}

\title{Newtonian Hydrodynamics with General Relativistic Pressure}
\author{Jai-chan Hwang}
\address{Department of Astronomy and Atmospheric Sciences,
         Kyungpook National University, Daegu 702-701, Republic of Korea}
\ead{jchan@knu.ac.kr}
\author{Hyerim Noh}
\address{Korea Astronomy and Space Science Institute,
         Daejeon 305-348, Republic of Korea}
\ead{hr@kasi.re.kr}


\begin{abstract}
We present the general relativistic pressure correction terms in
Newtonian hydrodynamic equations to the nonlinear order: these are equations (\ref{mass-conservation-Mink})-(\ref{Poisson-eq-Mink}). The
derivation is made in the zero-shear gauge based on the fully nonlinear formulation of cosmological perturbation in Einstein's gravity. The correction terms {\it differ} from many of the previously suggested forms in the literature based on hand-waving manners. We confirm our results by comparing with (i) the nonlinear perturbation theory, (ii) the first order post-Newtonian approximation, and (iii) the special relativistic limit, and by checking (iv) the consistency with full Einstein's equation.
\end{abstract}

\tableofcontents

%
%
\section{Introduction}

The general relativistic pressure corrections in Newtonian
hydrodynamic equations have attracted some attentions in the
cosmology literature. Previously the pressure correction terms were
guessed in hand-waving manner without properly based on relativistic
gravity
\cite{Whittaker-1935,McCrea-1951,Harrison-1965,Coles-Lucchin-2002}.
The proper study should begin with Einstein's gravity and take
Newtonian limit except for keeping the relativistic and gravitating
pressure terms. This is the aim of our present work.

Based on the formulation of fully nonlinear cosmological
perturbation theory in \cite{Hwang-Noh-2013-FNL}, recently we have
derived the exact Newtonian hydrodynamic equations as the infinite
speed-of-light limit (weak gravity, nonrelativistic speed, subhorizon, and negligible
pressure compared with the energy density) in two gauge conditions
\cite{Hwang-Noh-2013-Newtonian}: the zero-shear gauge and the uniform-expansion gauge. Here, by relaxing only the
condition on the pressure term we derive the Newtonian
hydrodynamic equations in the presence of relativistic and
gravitating pressure correction terms in the zero-shear gauge.
The cosmological hydrodynamic equations include the ordinary nonlinear hydrodynamic equations as a case.

Our main result is the Newtonian hydrodynamic equations
with the general relativistic pressure corrections. In the presence of relativistic pressure the mass conservation, momentum conservation, and the Poisson's equation, respectively, become:
\bea
   & & \dot {\widetilde \varrho}
         + \nabla \cdot
       \left[ \left( \widetilde \varrho
         + {\widetilde p \over c^2} \right) {\bf v} \right]
       = {2 \over c^2}
       {\bf v} \cdot \nabla \widetilde p,
   \label{mass-conservation-Mink} \\
   & & \dot {\bf v}
       + {\bf v} \cdot \nabla {\bf v}
       - \nabla U
       = - {1 \over \widetilde \varrho + \widetilde p/c^2}
       \left( \nabla \widetilde p
       + {\dot {\widetilde p} \over c^2} {\bf v} \right),
   \label{momentum-conservation-Mink} \\
   & & \Delta U
       = - 4 \pi G \widetilde \varrho,
   \label{Poisson-eq-Mink}
\eea
where $\widetilde \varrho$, $\widetilde p$, ${\bf v}$, and $U$ are the mass density (including the internal energy), the pressure, the velocity, and the gravitational potential, respectively.
In the context of flat Friedmann background the equations become equations (\ref{mass-conservation})-(\ref{Poisson-eq}).

The derived corrections are new in the sense that some of the pressure
corrections {\it differ} from the previously guessed forms in the literature based on pseudo-Newtonian manner: compare our equations
(\ref{mass-conservation-Mink})-(\ref{Poisson-eq-Mink}) with the ones
in equations (10), (11b) and (13) in
\cite{Harrison-1965}, equations (10.9.1a-c) in
\cite{Coles-Lucchin-2002}. Many of the subsequent works in the literature
were based on these {\it incorrect} equations. The comparison
reveals that the pressure term in the right-hand-side of equation
(\ref{mass-conservation-Mink}) was not recognized in the previous studies. For example, equations (10.9.1a-c) in a textbook \cite{Coles-Lucchin-2002} differ completely from ours in all three equations.
Based on semi-post-Newtonian treatment the author of \cite{Harko-2011} has
shown equation (\ref{momentum-conservation-Mink}) but failed to show
equation (\ref{mass-conservation-Mink}): see equations (23), (25) and (26) in \cite{Harko-2011}. Equation (\ref{momentum-conservation-Mink}) can be found in equation (2.10.16) of \cite{Weinberg-1972} based on the special relativistic hydrodynamics (thus without assuming nonrelativistic speed) in the absence of gravity.

More remarkably, notice the absence of pressure correction term in the Poisson's equation contrary to the common belief made in the literature \cite{Whittaker-1935,Harrison-1965,Coles-Lucchin-2002,Peacock-1999,Harko-2011}.
See comments below equation (\ref{Poisson-2}).

Our aim in this work is to derive equations
(\ref{mass-conservation-Mink})-(\ref{Poisson-eq-Mink}) from the fully
nonlinear cosmological perturbation equations in Einstein's gravity
(section \ref{sec:proof}). We will check the validity of our pressure correction terms by comparing our equations with the perturbation theory up to the third order (section \ref{sec:PT}) and with the first-order post-Newtonian hydrodynamic equations (section \ref{sec:PN}). For an easy comparison we present the fully nonlinear cosmological perturbation equations in Einstein's gravity in the Appendix.

%
%
\section{Notation}

We consider the scalar- and vector-type perturbations in a {\it
flat} background with the metric convention
\cite{Bardeen-1988,Hwang-Noh-2013-FNL} \bea
   & & ds^2 = - \left( 1 + 2 \alpha \right) c^2 d t^2
       - 2 \chi_i c d t d x^i
       + a^2 \left( 1 + 2 \varphi \right) \delta_{ij} d x^i d x^j,
   \label{metric-PT}
\eea where $a(t)$ is the cosmic scale factor; $\alpha$, $\varphi$ and $\chi_i$ are functions of
spacetime with arbitrary amplitudes; index of $\chi_i$ is raised and lowered by $\delta_{ij}$ as the metric. The spatial part of the metric is
simple because we already have taken the spatial gauge condition
without losing any generality to the fully nonlinear order
\cite{Bardeen-1988,Hwang-Noh-2013-FNL}, and have {\it ignored} the transverse-tracefree part of the metric perturbation.

We consider a fluid {\it without} anisotropic stress. The energy momentum
tensor is \bea
   \widetilde T_{ab}
       = \widetilde \varrho c^2 \widetilde u_a \widetilde u_b
       + \widetilde p \left( \widetilde g_{ab} + \widetilde u_a \widetilde u_b \right),
   \label{Tab}
\eea
where tildes indicate the
covariant quantities; $\widetilde u_a$ is the normalized fluid four-vector. We
set \bea
   \widetilde u_i \equiv a {v_i \over c},
\eea where the index of $v_i$ is raised and lowered by $\delta_{ij}$ as the metric. We introduce the fluid three-velocity $\widehat v_i$
measured by the Eulerian observer with the normal four-vector
$\widetilde n^c$. It is related to $v_i$ as \cite{Hwang-Noh-2013-FNL} \bea
   v_i \equiv \widehat \gamma \widehat v_i,
\eea with \bea
   \widehat \gamma
       \equiv \sqrt{ 1
       + {v^k v_k \over c^2 (1 + 2 \varphi)}}
       = {1 \over \sqrt{ 1
       - {\widehat v^k \widehat v_k \over c^2 (1 + 2 \varphi)}}},
\eea the Lorentz factor; the index of $\widehat v_i$ is raised and lowered by $\delta_{ij}$.

We can decompose $\chi_i$ and $\widehat v_i$ into the scalar- and
vector-type perturbations to the nonlinear order as
\cite{Hwang-Noh-2013-FNL} \bea
   \chi_i = c \chi_{,i} + a \Psi_i^{(v)}, \quad
       \widehat v_i \equiv - \widehat v_{,i} + \widehat v_i^{(v)},
\eea with $\Psi^{(v)i}_{\;\;\;\;\;\;,i} \equiv 0 \equiv
\widehat v^{(v)i}_{\;\;\;\;\;\;,i}$.  Dimensions are presented in \cite{Hwang-Noh-2013-FNL,Hwang-Noh-2013-Newtonian}.

The complete set of fully nonlinear perturbation equations without
taking the temporal gauge was derived in Equations (28)-(38) of \cite{Hwang-Noh-2013-FNL}. The basic equations in \cite{Hwang-Noh-2013-FNL} were presented using $v_i$ as the perturbed velocity. In the Appendix we present the same equations now using $\widehat v_i$ as the perturbed velocity; we recover $c$, and $\widetilde \varrho$ includes the internal energy; in explicit presence of the internal energy density we should replace \bea
   \widetilde \varrho
       \rightarrow \widetilde \varrho
       \left( 1 + {1 \over c^2} \widetilde \Pi \right),
   \label{internal-energy}
\eea where $\widetilde \varrho$ in the right-hand-side is the
rest-mass density \cite{Chandrasekhar-1965}.

Our basic set has seven equations presented in the Appendix: these are (i) the definition of $\kappa$ which is the perturbed part of trace of extrinsic curvature, (ii) the ADM energy constraint, (iii) the ADM momentum constraint, (iv) the trace of ADM propagation, (v) the tracefree ADM propagation, (vi) the covariant energy conservation, and (vii) the covariant momentum conservation. In the following we will call these equations using the names assigned above.

%
%
\section{Infinite speed-of-light limit except for pressure}
                                          \label{sec:limit}

In order to show the Newtonian limit, in \cite{Hwang-Noh-2013-Newtonian} we have taken the infinite speed-of-light limit.
Here we relax the nonrelativistic condition on
pressure, thus we do {\it not} assume $\widetilde p \ll \widetilde
\varrho c^2$; we also do {\it not} assume $\widetilde \Pi/c^2 \ll
1$. Thus, as the non-relativistic limit we consider the weak-gravity and the slow-motion limits \bea
   \alpha \ll 1, \quad
       \varphi \ll 1, \quad
       {\widehat v^k \widehat v_k \over c^2} \ll 1.
   \label{NL-limit}
\eea Under this slow-motion limit we have $v^i = \widehat v^i$. We identify \bea
   \alpha = - {1 \over c^2} U, \quad
       \varphi = {1 \over c^2} V, \quad
       \widehat v^k = {\bf v},
   \label{identification}
\eea where ${\bf v}$ is the perturbed Newtonian velocity; $U$ and
$V$ correspond to the Newtonian and the post-Newtonian perturbed
gravitational potentials, respectively \cite{Chandrasekhar-1965,Hwang-etal-2008}; later we will show $\varphi = - \alpha$ in our case, thus $V = U$. As
the subhorizon limit, we take the dimensionless quantity \bea
   {c^2 k^2 \over a^2 H^2} \gg 1,
   \label{SS-limit}
\eea where $k$ is the comoving wave-number with $\Delta = - k^2$; $H \equiv \dot a/a$; in the presence of the cosmological
constant $\Lambda$, we consider $H^2 \sim 8 \pi G \varrho$.

%
%
\section{Proof in the zero-shear gauge}
                                          \label{sec:proof}

In this section we will derive the following equations from the fully nonlinear cosmological perturbation equations in Einstein's gravity presented in the Appendix.
\bea
   & & \dot {\widetilde \varrho}
         + 3 {\dot a \over a} \left( \widetilde \varrho
         + {\widetilde p \over c^2} \right)
         + {1 \over a} \nabla \cdot
       \left[ \left( \widetilde \varrho
         + {\widetilde p \over c^2} \right) {\bf v} \right]
       = {1 \over c^2} {2 \over a}
       {\bf v} \cdot \nabla \widetilde p,
   \label{mass-conservation} \\
   & & \dot {\bf v} + {\dot a \over a} {\bf v}
       + {1 \over a} {\bf v} \cdot \nabla {\bf v}
       - {1 \over a} \nabla U
       = - {1 \over \widetilde \varrho + \widetilde p/c^2}
       \left( {1 \over a} \nabla \widetilde p
       + {\dot {\widetilde p} \over c^2} {\bf v} \right),
   \label{momentum-conservation} \\
   & & {\Delta \over a^2} U
       = - 4 \pi G \left( \widetilde \varrho - \varrho \right),
   \label{Poisson-eq}
\eea
where we decompose the mass density $\widetilde
\varrho$ and the pressure $\widetilde p$ into the background and
perturbation as \bea
   \widetilde \varrho = \varrho + \delta \varrho, \quad
       \widetilde p = p + \delta p.
\eea Equation (\ref{mass-conservation}) can be decomposed to the
background and perturbation orders as \bea
   & & \dot \varrho + 3 {\dot a \over a}
       \left( \varrho + {p \over c^2} \right) = 0,
   \\
   & & \dot \delta
       + 3 {\dot a \over a} {1 \over c^2}
       \left( {\delta p \over \varrho}
         - {p \over \varrho} \delta \right)
         + {1 \over a} \nabla \cdot
       \left[ \left( 1 + \delta \right) {\bf v} \right]
       = {1 \over c^2} {1 \over a \varrho} \left(
       {\bf v} \cdot \nabla \widetilde p
       - \widetilde p \nabla \cdot {\bf v} \right),
   \label{mass-conservation-2}
\eea where $\delta \equiv \delta \varrho/\varrho$.

By taking the infinite-speed-of-light limit, the above equations
properly reduce to the Newtonian hydrodynamic equations
\cite{Hwang-Noh-2013-Newtonian}; for the Newtonian derivation, see
sections 7 and 9 in \cite{Peebles-1980}.

In the Minkowski background, the background order
quantities become $a = 1$ and $\varrho = 0 = p$, 
thus $\delta \varrho = \widetilde \varrho$ and $\delta p =
\widetilde p$. Equations
(\ref{mass-conservation})-(\ref{Poisson-eq}) properly become equations (\ref{mass-conservation-Mink})-(\ref{Poisson-eq-Mink}).

Now, the derivation of equations (\ref{mass-conservation})-(\ref{Poisson-eq}) goes as the following.
We consider the zero-shear gauge ($\chi \equiv 0$). In \cite{Hwang-Noh-2013-Newtonian} we have shown that compared with
$\widehat v_i^{(v)}$, $\Psi_i^{(v)}$ is suppressed by the small-scale
condition in equation (\ref{SS-limit}). Thus, $\chi_i = c \chi_{,i}$,
while keeping $\widehat v_i^{(v)}$. Thus, in the subhorizon limit and in the zero-shear gauge we have \bea
   \chi_i = 0.
   \label{chi_i}
\eea

The tracefree ADM propagation equation gives \bea
   \varphi = - \alpha.
   \label{varphi-alpha}
\eea The ADM momentum constraint equation in the zero-shear gauge gives \bea
   \kappa = - {12 \pi G a \over c^2} \Delta^{-1}
       \nabla \cdot \left[
       \left( \widetilde \varrho
       + {\widetilde p \over c^2} \right) {\bf v} \right],
   \label{kappa-chi}
\eea where $\kappa$ is a perturbed part of the trace of extrinsic
curvature (equivalently, a perturbed part of the expansion scalar
of the normal-frame vector with a minus sign).

Using equations (\ref{varphi-alpha}) and (\ref{kappa-chi}), the covariant energy conservation and the covariant momentum conservation equations give the conservation equations in (\ref{mass-conservation}) and (\ref{momentum-conservation}), respectively.

The Poisson's equation in (\ref{Poisson-eq}) follows from either
the ADM energy constraint equation or the trace of ADM propagation equation. The derivations deserve special comments. Under our limits, the ADM energy constraint equation and the trace of ADM propagation equation, respectively, become \bea
   & &
       + c^2 {\Delta \over a^2} \varphi
       + 4 \pi G \delta \varrho
       \qquad \qquad \,
       =
       \quad\;\,
       - {\dot a \over a} \kappa,
   \label{Poisson-1} \\
   & &
       - c^2 {\Delta \over a^2} \alpha
       + 4 \pi G \left( \delta \varrho
       + 3 {\delta p \over c^2} \right)
       = \dot \kappa + 2 {\dot a \over a} \kappa.
   \label{Poisson-2}
\eea
Using equation (\ref{kappa-chi}) we can see the $\kappa$ terms in the right-hand-sides are negligible. However, it is essentially important to keep $\dot \kappa$ term. Using equations (\ref{mass-conservation}), (\ref{momentum-conservation}) and (\ref{kappa-chi}) we can show that the $\dot \kappa$ term becomes $12 \pi G \delta p/c^2$, thus exactly cancels the $12 \pi G \delta p/c^2$ term in the left-hand-side of equation (\ref{Poisson-2}). Thus, both equations (\ref{Poisson-1}) and (\ref{Poisson-2}) give equation (\ref{Poisson-eq}). The left-hand-side of equation (\ref{Poisson-2}), the Raychaudhuri equation, shows apparent presence of the pressure contribution to the gravity often emphasized in the literature \cite{Harrison-1965,Coles-Lucchin-2002,Harko-2011}. However, as we just showed $\dot \kappa$ term exactly cancels the pressure term in our limits. We can also check the above statements by using the linear perturbation solutions in the zero-shear gauge for an ideal fluid with constant $w \equiv p/(\varrho c^2)$ and $\delta p = w \delta \varrho c^2$; the complete subhorizon scale solutions in all fundamental gauges are presented in Table 9 of \cite{Hwang-1993-IF}\footnote{In the Table 9 of \cite{Hwang-1993-IF} we wish to correct an error by replacing $S^\prime$ to $x(S/x)^\prime$.}. For a similar analysis in the case of the post-Newtonian approximation, see below equation (\ref{Poisson-eq-PN}).

Here we would like to add further comments on this subtle issue.
Equations (\ref{Poisson-1}) and (\ref{Poisson-2}) corresponds to the $\widetilde G^{00}$ and $\widetilde R^{00}$ parts of Einstein's equation, respectively. In the presence of relativistic pressure we no longer have the conventional Newtonian limit argument; for example, in the presence of pressure, we no longer have $\widetilde G^{00} = 2 \widetilde R_{00}$ as is often available in the Newtonian limit argument used, for example, in equation (1.81) of \cite{Peacock-1999}. Notice that whereas $\widetilde G^{00}$-equation is a constraint equation, $\widetilde R^{00}$-equation gives a propagation equation. In the presence of the relativistic pressure the constraint equation gives a Poisson's equation without pressure whereas the propagation equation apparently have the pressure correction term: see equations (\ref{Poisson-1}) and (\ref{Poisson-2}). At this point, we have to carefully examine the time-derivative term ($\dot \kappa$-term in perturbation theory, and $\ddot U$-term in the post-Newtonian approximation) in the propagation equation, which turns out to exactly {\it cancel} the pressure term as we have explained in the previous paragraph.

For the proof of equations (\ref{mass-conservation})-(\ref{Poisson-eq}) we have used six equations in our basic set consisting of seven equations.
Thus, we still have one more equation (the definition of $\kappa$) in Einstein's gravity which needs to be checked for consistency. This demands a careful examination of the equations with an issue which will be resolved in the following.
In our limits, the definition of $\kappa$ equation leads to
\bea
   - \dot \varphi + H \alpha = {1 \over 3} \kappa.
\eea By na\"ively using equations (\ref{mass-conservation}), (\ref{momentum-conservation}), (\ref{varphi-alpha})-(\ref{Poisson-2}) this equation could lead to an inconsistency by a term $c^{-4} 8 \pi G a \Delta^{-1} ( {\bf v} \cdot \nabla \widetilde p)$ present in the left-hand-side. However, by using the identifications in equation (\ref{identification}) and equation (\ref{kappa-chi}), we notice that both sides of this equation are suppressed by the small-scale limit factor in equation (\ref{SS-limit}). Thus, as both sides are higher order in the small-scale expansion, the equation cannot be relied in our limits. For a similar issue addressed in the
context of proving the Newtonian limit, see section 5 in
\cite{Hwang-Noh-2013-Newtonian}.

This completes the proof of equations (\ref{mass-conservation})-(\ref{Poisson-eq}), thus consequently equations (\ref{mass-conservation-Mink})-(\ref{Poisson-eq-Mink}) as well.

%
%
\section{Confirmation from perturbation theory}
                                                      \label{sec:PT}

Using the gauge-ready form perturbation equations presented to the third order in section 4 of \cite{Hwang-Noh-2013-FNL}, we have checked the validity of equations (\ref{mass-conservation})-(\ref{Poisson-eq}) to the third order perturbation in the zero-shear gauge; this is natural (even a tautology), though, because in the previous section we already have proved these equations based on the fully nonlinear equations where the third-order perturbation equations are the simple consequence of those.

Here, as we have some issues to clarify, we would like to examine the linear order case more closely.
To the linear order equations
(\ref{mass-conservation})-(\ref{Poisson-eq}) give \bea
   & & \dot \delta
       + 3 {\dot a \over a} {1 \over c^2}
       \left( {\delta p \over \varrho}
         - {p \over \varrho} \delta \right)
         + {1 \over a} \left( 1 + {p \over \varrho c^2} \right)
         \nabla \cdot {\bf v}
       = 0,
   \label{mass-conservation-lin} \\
   & & \dot {\bf v} + {\dot a \over a} {\bf v}
       - {1 \over a} \nabla U
       = - {1 \over \varrho + p/c^2}
       \left( {1 \over a} \nabla \delta p
       + {\dot p \over c^2} {\bf v} \right),
   \label{momentum-conservation-lin} \\
   & & {\Delta \over a^2} U
       = - 4 \pi G \delta \varrho.
   \label{Poisson-eq-lin}
\eea Once again, we can check that these are the same as linear perturbation equations in the zero-shear gauge in the small-scale (subhorizon) limit.

From equations (\ref{mass-conservation-lin})-(\ref{Poisson-eq-lin}) we can derive \bea
   \fl \ddot \delta
       + \left( 2 + 3 {c_s^2 \over c^2} - 6 w \right) H \dot \delta
       + \Bigg[ 3 H {\partial \over \partial t} \left( {c_s^2 \over c^2} - w \right)
       + 3 \dot H \left( {c_s^2 \over c^2} - w \right)
       + 3 H^2 \left( {c_s^2 \over c^2} - w \right)
       \left( 2 - 3 w \right)
   \nonumber \\
   \fl \qquad
       - 4 \pi G \varrho \left( 1 + w \right) \Bigg] \delta
       = c_s^2 {\Delta \over a^2} \delta
       + {\Delta \over a^2} {e \over \varrho}
       - 3 H {\dot e \over \varrho c^2}
       - 4 \pi G \left( 7 - 3 w \right) {e \over c^2},
   \label{ddot-eq-linear}
\eea where \bea
   & & \delta p \equiv c_s^2 \delta \varrho + e, \quad
       c_s^2 \equiv {\dot p \over \dot \varrho}, \quad
       w \equiv {p \over \varrho c^2}.
\eea This equation is valid only in the subhorizon limit. Thus, applying the subhorizon limit condition, we have \bea
   \ddot \delta
       + \left( 2 + 3 {c_s^2 \over c^2} - 6 w \right) H \dot \delta
       = c_s^2 {\Delta \over a^2} \delta
       + {\Delta \over a^2} {e \over \varrho}.
   \label{delta-eq-CG-SS}
\eea This {\it coincides} with the density perturbation equation in
the comoving gauge ($v \equiv 0$) in the subhorizon limit
\cite{Nariai-1969,Bardeen-1980}, see equation (45) in
\cite{Hwang-Noh-1999}.

In the zero-pressure limit, equation
(\ref{ddot-eq-linear}) properly reduces to \bea
   \ddot \delta + 2 H \dot \delta - 4 \pi G \varrho \delta
       - c_s^2 {\Delta \over a^2} \delta = 0,
   \label{delta-eq-N}
\eea which is the well known equation in the comoving gauge ($v \equiv 0$) and the synchronous gauge ($\alpha \equiv 0$) available for a zero-pressure fluid
\cite{Lifshitz-1946,Nariai-1969}; we have ignored $c_s^2/c^2$, but
have kept the bare $c_s^2$ term, and similarly have kept $4 \pi G
\varrho$ term which was ignored in equation (\ref{delta-eq-CG-SS})
due to the small-scale limit in the presence of substantial pressure
with $c_s^2 \sim c^2$. Notice that equation (\ref{delta-eq-N}) is valid in the zero-shear gauge in the subhorizon limit, whereas the same equation is valid in all scales in the comoving gauge \cite{Nariai-1969,Bardeen-1980,Hwang-1993-IF,Hwang-Noh-1999}.

%
%
\section{Confirmation from post-Newtonian approximation}
                                                      \label{sec:PN}

The cosmological post-Newtonian (PN) approximation presented in
\cite{Hwang-etal-2008} can provide another confirmation of equations
(\ref{mass-conservation})-(\ref{Poisson-eq}). The mass conservation,
the momentum conservation, and the Poisson's equations valid to the
first PN (1PN) order without taking the (temporal PN) gauge
condition are presented in equations (114), (115) and (119) of
\cite{Hwang-etal-2008}. The correspondence between the two
(perturbation versus PN) approaches was studied in
\cite{Noh-Hwang-2012-PN,Noh-Hwang-2013-PN} with identifications \bea
   \fl
       \alpha = - {1 \over c^2} \left[ U
       - {1 \over c^2} \left( U^2 - 2 \Phi \right) \right], \quad
       \varphi = {1 \over c^2} V, \quad
       \kappa = - {1 \over c^2} \left( 3 {\dot a \over a} U
       + 3 \dot V
       + {1 \over a} P^k_{\;\;,k} \right),
   \nonumber \\
   \fl
       \chi_i = {1 \over c^3} a P_i, \quad
       v_i = \overline{v}_i
       + {1 \over c^2} \left[
       \overline{v}_i
       \left( {1 \over 2} \overline{v}^2 + U + 2 V \right)
       - P_i \right],
   \label{PT-PN}
\eea where the left- and right-hand-sides correspond to the
perturbation and the PN notations, respectively; we have $V = U$,
and $\overline{v}_i$ indicates the $v_i$ in \cite{Hwang-etal-2008}; $\overline v_i$ is the fluid coordinate three velocity introduced in \cite{Hwang-Noh-2013-FNL} as \bea
   {1 \over a} {\overline v^i \over c}
       \equiv {\widetilde u^i \over \widetilde u^0},
\eea where the index of $\overline v_i$ is raised and lowered by $\delta_{ij}$; in our metric convention in equation (\ref{metric-PT}), an index $0$ indicates $x^0 = ct$.
Compared with the Newtonian identifications in equation
(\ref{identification}), the exact 1PN identifications above look
more complicated; we have not taken the temporal gauge condition in
the 1PN notation. By strictly applying the conditions in equation
(\ref{NL-limit}), equations (114), (115) and (119) in
\cite{Hwang-etal-2008} give \bea
   & & \dot {\widetilde \varrho}
       + 3 {\dot a \over a} {\widetilde \varrho}
       + {1 \over a} \nabla \cdot
       \left( \widetilde \varrho {\bf v} \right)
       = {1 \over c^2} \left[ {1 \over a} {\bf v} \cdot \nabla \widetilde p
       - \left( 3 {\dot a \over a}
       + {1 \over a} \nabla \cdot {\bf v} \right) \widetilde p \right],
   \label{mass-conservation-PN} \\
   & & \dot {\bf v} + {\dot a \over a} {\bf v}
       + {1 \over a} {\bf v} \cdot \nabla {\bf v}
       - {1 \over a} \nabla U
       + {1 \over a \widetilde \varrho}
       \nabla \widetilde p
       = {1 \over c^2} \left(
       {\widetilde p \over a \widetilde \varrho^2}
       \nabla \widetilde p
       - {\dot {\widetilde p} \over \widetilde \varrho} {\bf v} \right),
   \label{momentum-conservation-PN} \\
   & & {\Delta \over a^2} U
       = - 4 \pi G \delta \varrho.
   \label{Poisson-eq-PN}
\eea These equations coincide {\it exactly} with equations (\ref{mass-conservation})-(\ref{Poisson-eq}) expanded to the 1PN order.

The absence of pressure correction term in equation (\ref{Poisson-eq-PN}) again may deserve special comments. It follows from equation (119) of \cite{Hwang-etal-2008}; in our notation \bea
   & & {\Delta \over a^2} U
       = - 4 \pi G \left[ \widetilde \varrho - \varrho
       + 3 {1 \over c^2} \left( \widetilde p - p \right) \right]
       - {1 \over c^2} \Bigg\{ 3 \ddot U + 9 {\dot a \over a} \dot U
       + 6 {\ddot a \over a} U
   \nonumber \\
   & & \qquad
       + 8 \pi G \widetilde \varrho \overline v^i \overline v_i
       + {1 \over a^2} \left[ 2 \Delta \Phi
       - 2 U \Delta U + \left( a P^i_{\;\;, i} \right)^\cdot \right] \Bigg\}.
   \label{Poisson-eq-PN-exact}
\eea In order to show equation (\ref{Poisson-eq-PN}) it is important to keep $\ddot U/c^2$ term in equation (\ref{Poisson-eq-PN-exact}) compared with $\Delta U/a^2$, whereas $(\dot a/a) \dot U/c^2$ and $(\ddot a/a) U/c^2$ terms are negligible due to the small-scale condition.
In the PN approximation the zero-shear gauge corresponds to the transverse-shear gauge taking $P^k_{\;\;\; ,k} \equiv 0$ \cite{Hwang-etal-2008}. The $\ddot U/c^2$ term in equation (\ref{Poisson-eq-PN-exact}) {\it cancels} exactly with the pressure correction term which otherwise could have contributed to $- 12 \pi G \delta p$ term in the right-hand side of equation (\ref{Poisson-eq-PN}).
In order to show the cancelation we need the momentum-constraint equation presented in equation (120) of \cite{Hwang-etal-2008} as
\bea
   & & 0 = {1 \over a^2} \left( P^k_{\;\;|ki}
       - \Delta P_i \right)
       - 16 \pi G \widetilde \varrho \overline{v}_i
       + {4 \over a} \left( \dot U + {\dot a \over a} U \right)_{,i}.
   \label{Mom-constr-PN}
\eea
This follows from the ADM momentum constraint equation \cite{Noh-Hwang-2013-PN}.
In our case we can keep $\dot U_{,i} = 4 \pi G a \widetilde \varrho \overline v_i$ and using equations (\ref{mass-conservation-PN}) and (\ref{momentum-conservation-PN}) we can show $\ddot U = - 4 \pi G (\widetilde p - p)$, thus the $\ddot U$ term cancels the pressure term in equation (\ref{Poisson-eq-PN-exact}) exactly.
Exactly similar analysis was made in the nonlinear perturbation approach: see comments made in the two paragraphs below equation (\ref{Poisson-2}), and the relation between $\kappa$ and $U$ in equation (\ref{PT-PN}).

%
%
\section{Discussion}
                                                      \label{sec:Discussion}

In this work we have derived Newtonian hydrodynamic equations in the
presence of relativistic pressure corrections: these are equations
(\ref{mass-conservation-Mink})-(\ref{Poisson-eq-Mink}) in general, and equations (\ref{mass-conservation})-(\ref{Poisson-eq}) in the cosmological context.
We have derived these from the fully nonlinear cosmological
perturbation equations in Einstein's gravity \cite{Hwang-Noh-2013-FNL}.
As our pressure correction terms differ from the commonly known ones often used in the literature\cite{Harrison-1965,Coles-Lucchin-2002,Harko-2011}, in sections \ref{sec:PT} and \ref{sec:PN}
we have confirmed these equations by comparing with (i) the perturbation theory to the third order and (ii) the cosmological 1PN
approximation. In section \ref{sec:proof} we also have (iii) checked the self consistency of our equations using the full set of equations in Einstein's gravity.

The referee, Dr. Donghui Jeong, has suggested yet another independent check of our two conservation equations by comparing them with the ones in the special relativistic limit available in the literature. Indeed, we can show that (iv) our conservation equations in (\ref{mass-conservation-Mink}) and (\ref{momentum-conservation-Mink}) correctly reproduce the equations in the special relativistic limit (thus ignoring gravity, but keeping the relativistic speed and pressure): these are equation (2.10.16) in \cite{Weinberg-1972} and equations (2.3) and (2.4) in \cite{Peacock-1999}; these equations are valid for the relativistic speed whereas we assume the nonrelativistic speed. This indicates that the previously suggested pressure correction terms in the literature \cite{Harrison-1965,Coles-Lucchin-2002,Harko-2011} even fail to have correct special relativistic limit.

Applications of our hydrodynamic equations (\ref{mass-conservation-Mink})-(\ref{Poisson-eq-Mink}) with general relativistic pressure are left for future studies.

%
%
\section*{Acknowledgments}
We wish to thank the referee, Dr. Donghui Jeong, for careful examination of the manuscript with numerous important suggestions including the case of special relativistic limit.
H.N.\ was supported by grant No.\ 2012 R1A1A2038497 from NRF. J.H.\
was supported by KRF Grant funded by the Korean Government
(KRF-2008-341-C00022).

%
%


%
%
\section*{Appendix: Fully nonlinear perturbation formulation}
                                             \label{sec:NL-eqs}

Here we summarize the complete set of fully nonlinear perturbation
equations without taking the temporal gauge using $\widehat v_i$ as the fluid three-velocity \cite{Hwang-Noh-2013-FNL}. We recovered $c$ and $\widetilde \varrho \equiv \widetilde \mu/c^2$ includes the internal energy density.

\noindent
Definition of $\kappa$: \bea
   \fl \kappa
       \equiv
       3 {\dot a \over a} \left( 1 - {1 \over {\cal N}} \right)
       - {1 \over {\cal N} (1 + 2 \varphi)}
       \left[ 3 \dot \varphi
       + {c \over a^2} \left( \chi^k_{\;\;,k}
       + {\chi^{k} \varphi_{,k} \over 1 + 2 \varphi} \right)
       \right].
   \label{eq1}
\eea
ADM energy constraint:
\bea
   \fl - {3 \over 2} \left( {\dot a^2 \over a^2}
       - {8 \pi G \over 3} \widetilde \varrho
       - {\Lambda c^2 \over 3} \right)
       + {\dot a \over a} \kappa
       + {c^2 \Delta \varphi \over a^2 (1 + 2 \varphi)^2}
   \nonumber \\
   \fl \qquad
       = {1 \over 6} \kappa^2
       - 4 \pi G \left( \widetilde \varrho + {\widetilde p \over c^2} \right)
       \left( \widehat \gamma^2 - 1 \right)
       + {3 \over 2} {c^2 \varphi^{,i} \varphi_{,i} \over a^2 (1 + 2 \varphi)^3}
       - {c^2 \over 4} \overline{K}^i_j \overline{K}^j_i.
   \label{eq2}
\eea
ADM momentum constraint:
\bea
   \fl {2 \over 3} \kappa_{,i}
       + {c \over 2 a^2 {\cal N} ( 1 + 2 \varphi )}
       \left( \Delta \chi_i
       + {1 \over 3} \chi^k_{\;\;,ik} \right)
       + 8 \pi G \left( \widetilde \varrho + {\widetilde p \over c^2} \right)
       a \widehat \gamma^2 {\widehat v_{i} \over c^2}
   \nonumber \\
   \fl \qquad
       =
       {c \over a^2 {\cal N} ( 1 + 2 \varphi)}
       \Bigg\{
       \left( {{\cal N}_{,j} \over {\cal N}}
       - {\varphi_{,j} \over 1 + 2 \varphi} \right)
       \left[ {1 \over 2} \left( \chi^{j}_{\;\;,i} + \chi_i^{\;,j} \right)
       - {1 \over 3} \delta^j_i \chi^k_{\;\;,k} \right]
   \nonumber \\
   \fl \qquad
       - {\varphi^{,j} \over (1 + 2 \varphi)^2}
       \left( \chi_{i} \varphi_{,j}
       + {1 \over 3} \chi_{j} \varphi_{,i} \right)
       + {{\cal N} \over 1 + 2 \varphi} \nabla_j
       \left[ {1 \over {\cal N}} \left(
       \chi^{j} \varphi_{,i}
       + \chi_{i} \varphi^{,j}
       - {2 \over 3} \delta^j_i \chi^{k} \varphi_{,k} \right) \right]
       \Bigg\}.
   \label{eq3}
\eea
Trace of ADM propagation:
\bea
   \fl - 3 {1 \over {\cal N}}
       \left( {\dot a \over a} \right)^{\displaystyle\cdot}
       - 3 {\dot a^2 \over a^2}
       - 4 \pi G \left( \widetilde \varrho + 3 {\widetilde p \over c^2} \right)
       + \Lambda c^2
       + {1 \over {\cal {\cal N}}} \dot \kappa
       + 2 {\dot a \over a} \kappa
       + {c^2 \Delta {\cal N} \over a^2 {\cal N} (1 + 2 \varphi)}
   \nonumber \\
   \fl \qquad
       = {1 \over 3} \kappa^2
       + 8 \pi G \left( \widetilde \varrho + {\widetilde p \over c^2} \right)
       \left( \widehat \gamma^2 - 1 \right)
       - {c \over a^2 {\cal N} (1 + 2 \varphi)} \left(
       \chi^{i} \kappa_{,i}
       + c {\varphi^{,i} {\cal N}_{,i} \over 1 + 2 \varphi} \right)
       + c^2 \overline{K}^i_j \overline{K}^j_i.
   \label{eq4}
\eea
Tracefree ADM propagation:
\bea
   \fl \left( {1 \over {\cal N}} {\partial \over \partial t}
       + 3 {\dot a \over a}
       - \kappa
       + {c \chi^{k} \over a^2 {\cal N} (1 + 2 \varphi)} \nabla_k \right)
       \Bigg\{ {c \over a^2 {\cal N} (1 + 2 \varphi)}
   \nonumber \\
   \fl \qquad
       \times
       \left[
       {1 \over 2} \left( \chi^i_{\;\;,j} + \chi_j^{\;,i} \right)
       - {1 \over 3} \delta^i_j \chi^k_{\;\;,k}
       - {1 \over 1 + 2 \varphi} \left( \chi^{i} \varphi_{,j}
       + \chi_{j} \varphi^{,i}
       - {2 \over 3} \delta^i_j \chi^{k} \varphi_{,k} \right)
       \right] \Bigg\}
   \nonumber \\
   \fl \qquad
       - {c^2 \over a^2 ( 1 + 2 \varphi)}
       \left[ {1 \over 1 + 2 \varphi}
       \left( \nabla^i \nabla_j - {1 \over 3} \delta^i_j \Delta \right) \varphi
       + {1 \over {\cal N}}
       \left( \nabla^i \nabla_j - {1 \over 3} \delta^i_j \Delta \right) {\cal N} \right]
   \nonumber \\
   \fl \qquad
       =
       8 \pi G \left( \widetilde \varrho + {\widetilde p \over c^2} \right)
       \left[ {\widehat \gamma^2 \widehat v^i \widehat v_j \over c^2 (1 + 2 \varphi)}
       - {1 \over 3} \delta^i_j \left( \widehat \gamma^2 - 1 \right)
       \right]
       + {c^2 \over a^4 {\cal N}^2 (1 + 2 \varphi)^2}
   \nonumber \\
   \fl \qquad
       \times
       \Bigg[
       {1 \over 2} \left( \chi^{i,k} \chi_{j,k}
       - \chi_{k,j} \chi^{k,i} \right)
       + {1 \over 1 + 2 \varphi} \left(
       \chi^{k,i} \chi_k \varphi_{,j}
       - \chi^{i,k} \chi_j \varphi_{,k}
       + \chi_{k,j} \chi^k \varphi^{,i}
       - \chi_{j,k} \chi^i \varphi^{,k} \right)
   \nonumber \\
   \fl \qquad
       + {2 \over (1 + 2 \varphi)^2} \left(
       \chi^{i} \chi_{j} \varphi^{,k} \varphi_{,k}
       - \chi^{k} \chi_{k} \varphi^{,i} \varphi_{,j} \right) \Bigg]
       - {c^2 \over a^2 (1 + 2 \varphi)^2}
   \nonumber \\
   \fl \qquad
       \times
       \Bigg[ {3 \over 1 + 2 \varphi}
       \left( \varphi^{,i} \varphi_{,j}
       - {1 \over 3} \delta^i_j \varphi^{,k} \varphi_{,k} \right)
       + {1 \over {\cal N}} \left(
       \varphi^{,i} {\cal N}_{,j}
       + \varphi_{,j} {\cal N}^{,i}
       - {2 \over 3} \delta^i_j \varphi^{,k} {\cal N}_{,k} \right) \Bigg].
   \label{eq5}
\eea
Covariant energy conservation:
\bea
   \fl
       \left[ {\partial \over \partial t}
       + {1 \over a ( 1 + 2 \varphi )} \left( {\cal N} \widehat v^k
       + {c \over a} \chi^k \right) \nabla_k \right] \widetilde \varrho
       + \left( \widetilde \varrho + {\widetilde p \over c^2} \right)
       \Bigg\{
       {\cal N} \left( 3 {\dot a \over a} - \kappa \right)
   \nonumber \\
   \fl \qquad
       +
       {({\cal N} \widehat v^k)_{,k} \over a (1 + 2 \varphi)}
       + {{\cal N} \widehat v^k \varphi_{,k} \over a (1 + 2 \varphi)^2}
       + {1 \over \widehat \gamma}
       \left[ {\partial \over \partial t}
       + {1 \over a ( 1 + 2 \varphi )} \left( {\cal N} \widehat v^k
       + {c \over a} \chi^k \right) \nabla_k \right] \widehat \gamma \Bigg\}
       = 0.
   \label{eq6}
\eea
Covariant momentum conservation:
\bea
   \fl {1 \over a \widehat \gamma}
       \left[ {\partial \over \partial t}
       + {1 \over a ( 1 + 2 \varphi )} \left( {\cal N} \widehat v^k
       + {c \over a} \chi^k \right) \nabla_k \right]
       \left( a \widehat \gamma \widehat v_i \right)
       + \widehat v^k \nabla_i \left( {c \chi_k \over a^2 ( 1 + 2
       \varphi)} \right)
   \nonumber \\
   \fl \qquad
       + {c^2 \over a} {\cal N}_{,i}
       - \left( 1 - {1 \over \widehat \gamma^2} \right) {c^2 {\cal N}
       \varphi_{,i} \over a (1 + 2 \varphi)}
   \nonumber \\
   \fl \qquad
       + {1 \over \widetilde \varrho + {\widetilde p \over c^2}}
       \left\{
       {{\cal N} \over a \widehat \gamma^2} \widetilde p_{,i}
       + {\widehat v_i \over c^2}
       \left[ {\partial \over \partial t}
       + {1 \over a ( 1 + 2 \varphi )} \left( {\cal N} \widehat v^k
       + {c \over a} \chi^k \right) \nabla_k \right] \widetilde p \right\}
       = 0,
   \label{eq7}
\eea
where \bea
   \fl {\cal N} \equiv \sqrt{ 1 + 2 \alpha
       + {\chi^k \chi_k \over a^2 ( 1 + 2 \varphi )}}, \quad
       \overline{K}^i_j \overline{K}^j_i
       = {1 \over a^4 {\cal N}^2 (1 + 2 \varphi)^2}
       \Bigg\{
       {1 \over 2} \chi^{i,j} \left( \chi_{i,j} + \chi_{j,i} \right)
       - {1 \over 3} \chi^i_{\;\;,i} \chi^j_{\;\;,j}
   \nonumber \\
   \fl - {4 \over 1 + 2 \varphi} \left[
       {1 \over 2} \chi^i \varphi^{,j} \left(
       \chi_{i,j} + \chi_{j,i} \right)
       - {1 \over 3} \chi^i_{\;\;,i} \chi^j \varphi_{,j} \right]
       + {2 \over (1 + 2 \varphi)^2} \left(
       \chi^{i} \chi_{i} \varphi^{,j} \varphi_{,j}
       + {1 \over 3} \chi^i \chi^j \varphi_{,i} \varphi_{,j} \right) \Bigg\}.
   \nonumber \\
   \label{K-bar-eq}
\eea

\end{document}